# Current Crowding in a High-Efficiency Black Phosphorus Light-Emitting Diode Using a Reflective Back Contact

*Julien Brodeur, Éloïse Rahier, Mathieu Chartray-Pronovost, Étienne Robert, Oussama Moutanabbir, Stéphane Kéna-Cohen*[*]

Department of Engineering Physics, Polytechnique Montréal

We demonstrate a high-performance mid-infrared (MIR) light-emitting diode (LED) based on a black phosphorus (b-P)/n-MoS$_2$ heterojunction. A gold back contact combined with a rhenium-doped n-type MoS$_2$ layer is used to enhance light extraction. The device shows a MIR peak external quantum efficiency (EQE) of (1.6 ± 0.2) % at room temperature and a record (7.0 ± 0.5) % EQE at 77 K, with a maximum radiant power density of (108 ± 8) W/cm$^2$. Finite-element simulations highlight the importance of phonon-assisted band-to-band tunneling under reverse bias and the influence of carrier velocity saturation under forward bias. The simulations also reveal that the high ideality factors extracted from the current-voltage characteristic are due to current crowding at the heterojunction and a consequence of the device geometry. These findings establish a new high-performance b-P LED architecture and provide crucial insights into the physics of MIR sources based on 2D materials.

Black phosphorus (b-P) has emerged as a promising material for mid-infrared (MIR) light-emitting diodes (LEDs), offering a compelling alternative to conventional epitaxially-grown devices[1]. Black phosphorus possesses technologically interesting optoelectronic properties such as a high carrier mobility[2], a direct and tunable bandgap across a significant portion of the MIR spectrum[3,4], a high photoluminescence quantum yield (PLQY), and a low Auger recombination coefficient[5]. Furthermore, the van der Waals nature of b-P allows for simple device architectures compared to conventional bulk alloys, quantum well and superlattice structures[6] as well as flexibility in device design due to the absence of any lattice matching constraints[7]. Despite these advantages, environmental degradation of b-P poses a challenge, particularly under high current injection conditions. Encapsulation, however, has been shown to be effective in mitigating degradation at moderate drive currents[8,9]. Finally, the scalability of b-P continues to be an obstacle for commercial production, but promising solutions are emerging based on chemical vapor deposition[10], pulsed laser deposition[11,12] and solution processing with b-P inks.[13,14,15,16]

In this work, we demonstrate a b-P/n-MoS$_2$ LED architecture where the bottom gold contact also plays the role of a back reflector. The rhenium-doped n-type MoS$_2$ flake acts as both a low-resistivity electron conducting layer and an optical spacer. At room temperature, our device shows a peak external quantum efficiency (EQE) of (1.6 ± 0.2) % and a radiant power density (13 ± 1) W/cm$^2$, which is on par with the best devices reported to date. At liquid nitrogen temperatures, we find an EQE of (7.0 ± 0.5) % and a radiant power density (108 ± 8) W/cm$^2$, which are the highest values ever observed in a b-P LED. To gain further insight into the device physics, we perform finite-element simulations of the current-voltage characteristic. Our simulations indicate that the forward bias current is dominated by carrier recombination and in the carrier velocity saturation regime, while the reverse bias current is well described by a phonon-assisted band-to-band tunneling mechanism. Importantly, our simulations also provide insights into the non-ideal diode behavior, revealing that current crowding at the heterojunction, due to the device geometry, contributes to the high ideality factors observed under forward bias.

Following the first b-P LED report,[17] nearly all of the reported device structures have been composed of anisotype heterojunctions, typically with naturally n-type $MoS_2$ as the electron transporting component and b-P serving as the p-type component.[17,18,19] One challenge pointed out early on[17] is that fabrication on $SiO_2$/Si or Si leads to low light extraction efficiency to air and EQEs well below 1%.[19,20,21] This was followed by the realization of devices combining metallic back mirrors with an optimized $Al_2O_3$ spacer, which dramatically improves light extraction and has been successful in demonstrating EQEs of >1%.[9,22]

Our LED structure fabricated on a $SiO_2$ (90 nm)/Si substrate is illustrated in Figure 1(a). It directly uses the cathode as a back reflector, while a rhenium-doped ($N_D \sim 5 \times 10^{18}$ cm$^{-3}$) n-type $MoS_2$ layer (2D Semiconductors) serves as the spacer. Deliberate doping of the chemical vapor transport-grown crystals allows for thick $MoS_2$ layers to be used, without suffering from resistive loss. Vertical injection is also beneficial to avoid current crowding as will be described below. Figure 1(b) shows a TEM image of the complete b-P/$MoS_2$/Au stack. Note that the oxide layer seen at the b-P/MoS2 interface is due to accidental delamination during sample preparation for TEM and is absent in other identically prepared LEDs as shown in Supporting Information Figure S1.

The layer thicknesses were carefully chosen to provide for optimized light extraction. The EQE of an LED can be expressed as:

$$\eta_{EQE} = \gamma \frac{q_{pl}}{1 - q_{pl}(1 - F)} F \eta_{out} \qquad (1)$$

where $\gamma$ is the injection efficiency defined as the fraction of injected carriers that recombine in the active region of the device, $q_{pl}$ is the b-P PLQY, $F$ is the Purcell Factor and $\eta_{out}$ is the outcoupling efficiency. For a low PLQY material, such as b-P, and modest Purcell Factors, Equation (1) reduces to:

$$\eta_{EQE} \approx \gamma q_{pl} F \eta_{out} \qquad (2)$$

indicating that the EQE is proportional to $F\eta_{out}$, the net power radiated in the viewing direction, as compared to the total power radiated by a dipole in a homogeneous environment. In Figure 1(c), we show the calculated emission enhancement $F\eta_{out}$ for a b-P/n-MoS$_2$/Au stack as a function and n-MoS$_2$ and b-P thicknesses $t_1$ and $t_2$.[23,24] For the calculation, we assume a purely in-plane dipole at the heterojunction, aligned along the armchair direction of b-P. Isotropic refractive indices were used for both b-P and MoS$_2$.[25,26] The inclusion of anisotropy was not found to significantly affect the enhancement factor. From Figure 1(c), we find $F\eta_{out}$ ~0.4 over a range of b-P and MoS$_2$ thicknesses. The fabricated device thicknesses are indicated by a dashed line on the Figure. For these thicknesses, optical losses are primarily due to radiation into waveguided modes within the high-index MoS$_2$ layer (see Supporting Information Figure S2). Moreover, the Purcell factor for this structure is $F \sim 2.5$, which indicates that the enhancement comes from increases in both the radiative decay rate of b-P and the outcoupling efficiency of the structure.

To confirm the enhanced outcoupling efficiency of the optimized stack, Figure 1(d) compares the measured room-temperature PL of b-P over the MoS$_2$/Au region of the LED to that of the b-P region directly above the SiO$_2$/Si substrate. We find a nine-fold increase in PL intensity from the b-P region on MoS$_2$/Au, as compared to the SiO$_2$/Si region. This can be explained by a combination of increased absorption in b-P, due to reflection of the pump laser by the mirror and increased $F\eta_{out}$ for this structure. Importantly, the calculated far-field radiation profiles (Supporting Information Figure S3) predict approximately Lambertian emission from both regions, indicating that there is no significant difference in light collection efficiency by the objective. PL imaging comparing the signal from a b-P/MoS$_2$/Au stack and the b-P/Au region of a similar device shows that the b-P emission is completely quenched by the Au contact in the absence of a spacer layer (Supporting Information Figure S4).

Figure 2(a) shows the current-voltage (IV) characteristics of the device, as a function of temperature. The reverse bias current deviates strongly from ideal diode behavior. Transport in b-P/MoS$_2$ LEDs was previously postulated to be band-to-band driven under reverse bias and thermionic under forward bias.[17] To better understand the forward bias behavior, the data was fit using the Shockley model, including a series resistance:

$$I(V) = I_s \left[ \exp\left(\frac{V - R_s I(V)}{nV_t}\right) - 1 \right] \qquad (3)$$

where $I_s$ is the reverse saturation current, $n$ is the diode ideality factor and $R_s$ is the series resistance, and the fit parameters are shown in Figure 2(b)(c) (see Figure S4 for the fits). We find very large ideality factors, $n > 20$, showing that the forward bias current also deviates from ideal diode behavior. Fit parameters obtained from the simulated I-V characteristics using finite element modelling are also shown in Figure 2(b)(c). These simulations will be discussed further below.

When examining the temperature dependence, we find that the increase in forward bias current at low temperature cannot be attributed to an increase in the b-P mobility because the reverse bias current is nearly temperature-independent and the IV curves only diverge at high forward bias. This was further confirmed by measuring the temperature-dependent IV characteristics of a 70 nm thick b-P resistor, which showed negligible variation in resistivity with temperature (see Figure S5). From the resistor IV characteristic and a mobility value of $(400 \pm 20)$ cm$^2$/Vs taken from the literature,[27] we can estimate the natural p-type doping of b-P to be $(8.5 \pm 0.5) \times 10^{17}$ cm$^{-3}$. Moreover, under reverse bias, the current increases almost linearly with voltage, exhibiting minimal temperature dependence. This is consistent with the previously reported band-to-band tunneling of b-P valence band electrons to the MoS$_2$ conduction band[28,29].

The temperature-dependent electroluminescence (EL) spectra were measured using Fourier Transform Infrared Spectroscopy (FTIR) and standard lock-in techniques, while driving the device with 400 µA square current pulses at 2 kHz (Figure 3(a)). The EL maximum redshifts from 3.8 µm at room temperature to 4.3 µm at 77K, accompanied by a five-fold increase in intensity. This enhancement is attributed to a reduction in non-radiative recombination at low temperature, consistent with previous findings[30] and our measured temperature-dependent PL of the same b-P flake (Supporting Information Figure S7). Even at 77 K, excitonic signatures are absent from the PL and EL spectra. The emission of b-P appears to be dominated by free carrier recombination in agreement with the weak excitonic binding energy of bulk b-P.[31] Figure 3(b) shows the measured EQE of the b-P LED as a function of the injected current density (see below for the device area). Details of the EQE calculation can be found in section 10 of the Supporting Information. The b-P LED achieves a peak EQE of approximately $(1.6 \pm 0.2)$ % at room temperature and a record $(7.0 \pm 0.5)$ % at 77 K. Notably, the device shows relatively low efficiency droop between the peak EQE and that at 4 kA/cm$^2$, varying from 5 % at 77 K to 37 % at room temperature. The efficiency droop is attributed to Auger recombination given the range of current densities involved.[22] The reduced efficiency droop at 77 K follows from the expected increase of the Auger lifetime at low temperature.[5] For reference, a table comparing the EQE of various MIR LEDs is shown in Table 2 of the Supporting Information. Figure 3(c) shows the LED radiant power density as a function of current density and temperature. We measured the devices up to a maximum forward bias of 15 V, which corresponded to a radiant power density of (13 ± 1) W/cm$^2$ at room temperature and (108 ± 8) W/cm$^2$ at 77 K.

To better understand the origin of the non-ideal behavior in the current-voltage characteristics of our LED, we performed 2D finite-element simulations of our device structure. Figure 4(a) shows the simulated equilibrium band diagram of the b-P/n-MoS$_2$ heterojunction. The band alignment reveals a near broken gap heterojunction with a space charge region less than 40 nm wide. Furthermore, the band alignment indicates that a small reverse bias of 0.1 V is enough to align the b-P valence and MoS$_2$ conduction bands leading to a high probability of band-to-band tunneling. Since the MoS$_2$ conduction band minimum is not located at the Γ point[32], the electronic transition from the b-P valence to MoS$_2$ conduction band is indirect and must involve phonon scattering to satisfy momentum conservation.

Our finite element model incorporates thermionic emission at the heterointerface, along with radiative, Shockley-Read-Hall (SRH), and Auger recombination within the b-P layer. The Schenk model[33] was used for phonon-assisted band-to-band tunneling in reverse bias and Caughey-Thomas carrier velocity saturation was also included. An isotropic mobility model was used as adding anisotropy does not significantly affect the simulation results due to the large geometric aspect ratio of the device as shown in Figure S8(b). Details of the simulation parameters can be found in section 7 of the Supporting Information. As seen in Figure 4 (b), the finite element model qualitatively reproduces the experimental IV curves of Figure 2 under both forward and reverse bias. While the simulated current is lower than the experimental values, this discrepancy could be due to an underestimation of the b-P dopant density due to the contact resistance of the b-P resistor (Supporting Information Figure S6). The simulated band diagram of the b-P/n-MoS$_2$ heterojunction at a 5 V in Supporting Information Figure S8 (a) shows a vanishing barrier for electrons in the conduction band suggesting that the forward bias current is not dominated by thermionic emission of electrons from MoS$_2$ to b-P as initially claimed.[17] As expected, removing thermionic emission of carriers at the heterointerface from the finite element model has a negligible impact on the resulting I-V curves in forward bias as shown in Supporting Information Figure S8 (b).

Figure 2 (a) and (b) also show the extracted ideality factor and series resistance as a function of temperature obtained by fitting the simulated IV curves with Eq. (3). The experimental room-temperature ideality factor was found to be approximately $n = 27$, increasing linearly to $n = 111$ at 77 K. Importantly, the simulated room-temperature ideality factor ($n \approx 23$) closely matches the experimental value and increases monotonically to ~90 at 77 K. Velocity saturation explains the observed temperature dependence of the current under forward bias. The voltage onset for velocity saturation increases at lower temperatures, due to the reduced carrier thermal velocity, which explains the series resistance trend in Figure 2(c).

To understand the origin of the high ideality factors, we can first consider a 1D version of the device using the same physical model (Figure 5(a) inset). The b-P and MoS$_2$ thicknesses and widths are adjusted to match the simulated 2D device series resistance. Figure 5(a) compares the calculated room-temperature forward bias current of the 1D device to that of the 2D structure. The 1D device shows an ideality factor (n ≈ 2) consistent with an ideal heterojunction, where multiple transport mechanisms (diffusion, recombination and high level injection) contribute to the current[34]. The Shockley model fit parameters for the 1D simulation are shown in Supporting Information Table S1. This result suggests that the high ideality factors observed in the 2D device are a consequence of its geometry. A contour map of the y-component of the room temperature current density, $J_y$, at 15 V is shown in Figure 5(b). We clearly observe significant current crowding at the b-P edge near the cathode, with the current density decaying exponentially along the junction. This non-uniform current distribution can also explain the device behaviour at low temperature. If the carrier mobility is temperature-independent, the diffusion coefficient will decrease at low temperatures following the Einstein relation $D(T) = \mu k_B T/e$ with $k_B$, the Boltzmann constant and $e$ the electron charge. Lateral diffusion of carriers in the b-P layer is thus reduced which exacerbates the current crowing effect. In the 1D limit, the current density is uniform, and the ideality factor approaches the expected value. It is crucial to note that the local current density cannot be obtained from the total current divided by the flake overlap area, as is often assumed for similar devices. The calculated current distribution in Figure 5(b) shows that more than 95 % of the total current crosses the junction within the first 6 μm along its length. While the flake overlap area is approximately 550 μm$^2$, we estimate the device active area to be closer to 120 μm$^2$ and this value was used to calculate the current and power densities reported in Figure 3. To unambiguously confirm the importance of current crowding, we have directly imaged the EL from a similar b-P LED using a MIR camera (Supporting Information Figure S10). The measurement shows that the EL is concentrated withing a few micrometers of the b-P edge confirming that the device active area is much smaller than the geometric flake overlap region. As further proof of the current crowding effect, Figure S9 compares the room-temperature EQE curve of Figure 3(b) with the PLQY of the b-P flake at equivalent carrier generation rates by assuming uniform current injection through the device area defined by the flake overlap. We find that the EQE roll-off due to Auger recombination occurs at a generation rate where the PLQY stays constant, which supports the fact that the current is flowing through a region much smaller than the geometric flake overlap. We suspect that a reduction in inhomogeneous current injection is also the source of the improvement in device performance of Ref. [22] when indium tin oxide is used to ensure vertical injection into $MoS_2$. That geometry, however, will still lead to current crowding at the (bottom) b-P edge. While large ideality factors in heterojunctions are often associated with increased non-radiative recombination due to the presence of interface traps,[35] this is not the case here. The non-ideal behavior is purely a result of current crowding and would disappear in a vertical device. Note also that the reported EQE is independent of the choice of device area and that the only consequence of using the "corrected" active area is to shift the injected current density (x-axis) of Figure 3(b).

In summary, we have demonstrated a simplified yet highly effective b-P LED structure, leveraging a gold back contact for enhanced MIR light extraction. Crucially, we have shown that a single n-doped $MoS_2$ flake can serve dual roles as both the n-type component of the p-n junction and a high-quality optical spacer to optimize light extraction. Our optimized b-P LED shows a peak EQE of $(1.6 \pm 0.2)$ % and a radiant power density of $(13 \pm 1)$ W/cm$^2$ at room temperature, which increases to $(7.0 \pm 0.5)$ % and $(108 \pm 8)$ W/cm$^2$, respectively, at 77 K, with minimal efficiency roll-off. Furthermore, we reveal the origin of the device IV characteristic, demonstrating through finite-element simulations the important roles of phonon-assisted band-to-band tunneling under reverse bias and diffusion of minority carrier combined with carrier velocity saturation in forward bias. Our simulations notably highlight the effect of current crowding on the non-ideal behavior of the b-P LED, which is a direct consequence of the 2D geometry. This finding underscores the importance of distinguishing between the physical device area and active area when reporting current densities or extracting physical parameters, such as Auger coefficients, from such devices. The forward bias ideality factor can be used to gauge the importance of current crowding in a given device structure. Our results not only establish a high-performance b-P LED architecture but also provide critical insights into the device physics, paving the way for further advancements in MIR light sources based on 2D materials. Indeed, current crowding could be completely mitigated through the addition of a transparent contact on top of b-P, such as graphene or a transparent conducting oxide. Such a device would show ideal diode behavior in forward bias, uniform MIR emission over the physical b-P/$MoS_2$ overlap area and a higher radiant power. Further improvement of the b-P LED EQE could be achieved with a double heterojunction using a low electron affinity material such as $WSe_2$ acting as an electron blocking layer to reach unity carrier injection efficiency.

METHODS

**Device fabrication**

The b-P LED was fabricated on a 90 nm $SiO_2$/Si substrate with gold contacts pre-patterned using standard photolithography. The 155 nm-thick n-type $MoS_2$ and 75 nm b-P flakes were mechanically exfoliated and stacked onto the gold contacts using a dry transfer technique. Flake thicknesses were determined via white light micro-reflectivity measurements during fabrication. Following this, a 1 nm Al layer was thermally evaporated onto the device and exposed to air to form a 1 nm $Al_2O_3$ encapsulation layer. The device was then annealed at 200°C for 1 hour on a hot plate to passivate native defects in b-P[36]. All fabrication steps other than oxidation were performed under a nitrogen atmosphere with oxygen and water vapor levels below 0.1 ppm.

**TEM measurement**

Thin specimens suitable for cross-sectional transmission electron microscopy (XTEM) observations were prepared by focused ion beam (FIB) using a FEI Helios NanoLab 600 operating with a Ga ion beam. The final thinning steps were performed at 2kV to minimize damage. TEM imaging was performed on a C-FEG JEOL JEM-Fx2000 operating at 200 kV.

**PL and EL measurements**

Both PL and EL measurements were performed under vacuum using a cryogenic stage (Linkam). For PL measurements, a 808 nm CW laser diode with a 10 μm spot diameter was focused onto the sample. The pump laser signal was modulated at 1 kHz using a mechanical chopper. For EL measurements, the device was driven with a 2 kHz square-wave voltage pulse train. In both cases, the emitted MIR signal was collected using a 20X reflective objective and directed to a FTIR spectrometer (Bruker Vertex V70) operating in step scan mode, equipped with an InSb photodetector. The detector output was then sent to a lock-in amplifier and the extracted signal was fed to the FTIR analog-to-digital converter.

**Finite-element modeling**

Synopsys Sentaurus was used for calculating the current-voltage characteristics of the device. The meshing size used was 200 nm along the junction (x-direction) and 10 nm perpendicular to the junction (y-direction) with an interface meshing density of 1 nm. The default drift-diffusion transport model was used for the simulation. The model also includes Fermi-statistics for carriers, thermionic emission at the heterointerface, recombination of carriers on b-P (radiative, SRH and Auger) and the Schenk model[33] for phonon-assisted band-to-band tunneling. The Caughey-Thomas model[37] for the high-field velocity saturation of carriers was also included in the model.

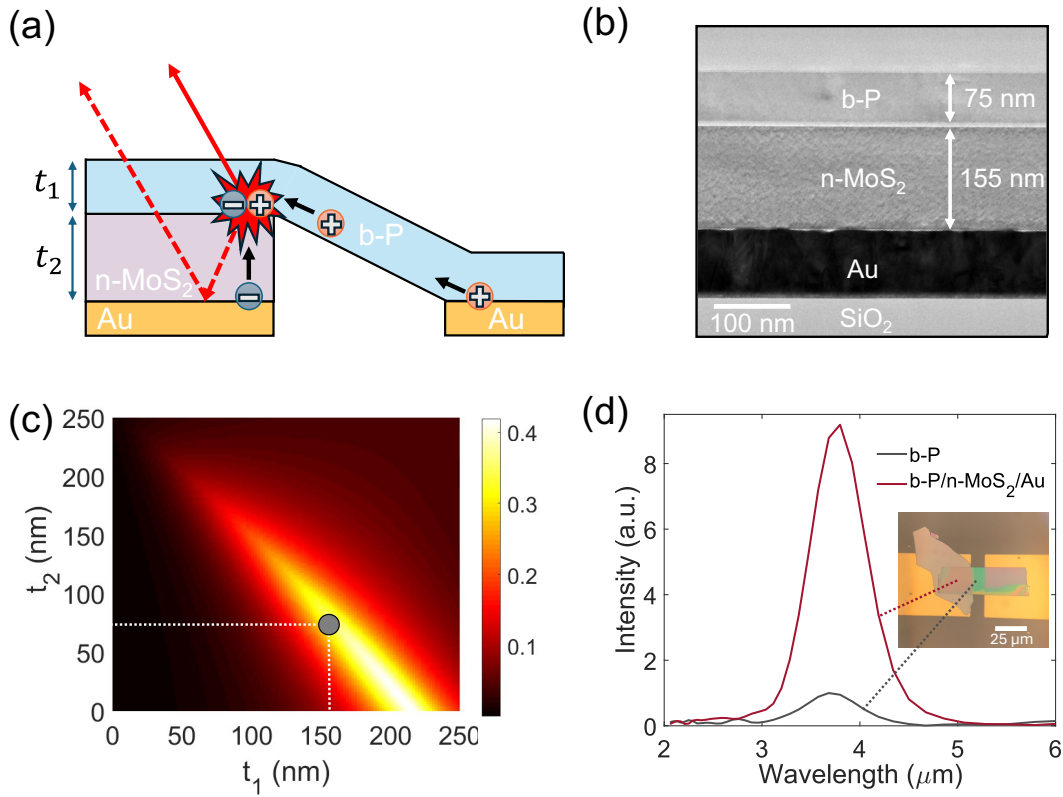

**Figure 1**: (a) Schematic of the LED architecture, where $t_1$ and $t_2$ are the n-MoS$_2$ and b-P thicknesses, respectively. (b) TEM image of the b-P/MoS$_2$/Au structure. (c) Calculated emission enhancement of the net power radiated in the viewing direction ($F\eta_{out}$), as a function of $t_1$ and $t_2$, assuming an in-plane transition dipole moment along the armchair direction. (d) Normalized room-temperature PL spectra of the 75 nm b-P flake on the MoS$_2$ (155 nm)/Au emission enhancing

structure and above the bare SiO$_2$ (90nm)/Si substrate. The spectra are obtained by pumping with an 808 nm laser diode at 10 mW. Inset: optical image of the fabricated b-P LED.

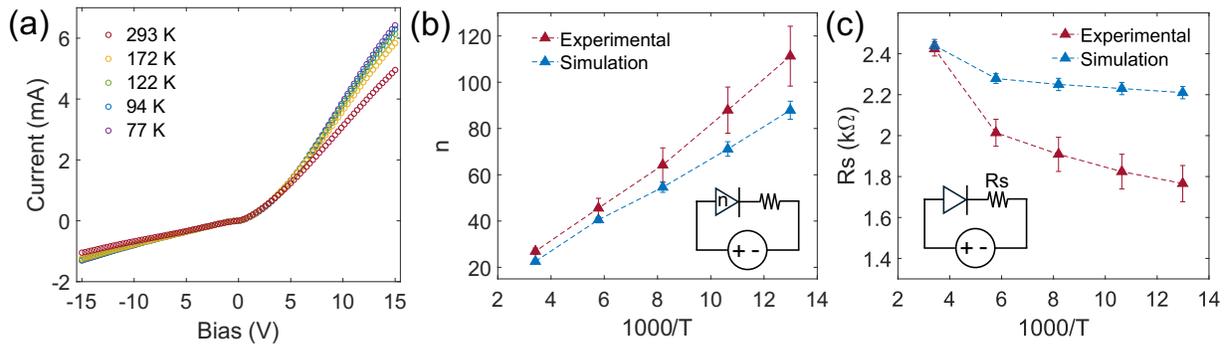

**Figure 2**: (a) IV characteristics of the b-P LED as a function of temperature. Ideality factors (b) and series resistance (c) extracted from the fits as a function of temperature for the experimental and simulated I-V curves.

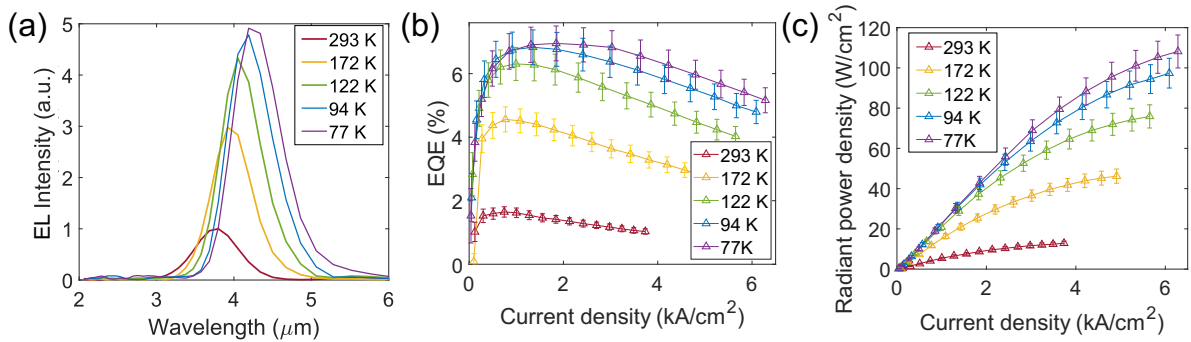

**Figure 3**: (a) Temperature dependent EL spectra normalized to the peak intensity at room temperature of the b-P LED operated using a 400 µA square pulse at 2 kHz. (b) External quantum efficiency (EQE) as a function of current density at different temperatures. (c) Electroluminescence (EL) radiant power density as a function of current density at different temperatures.

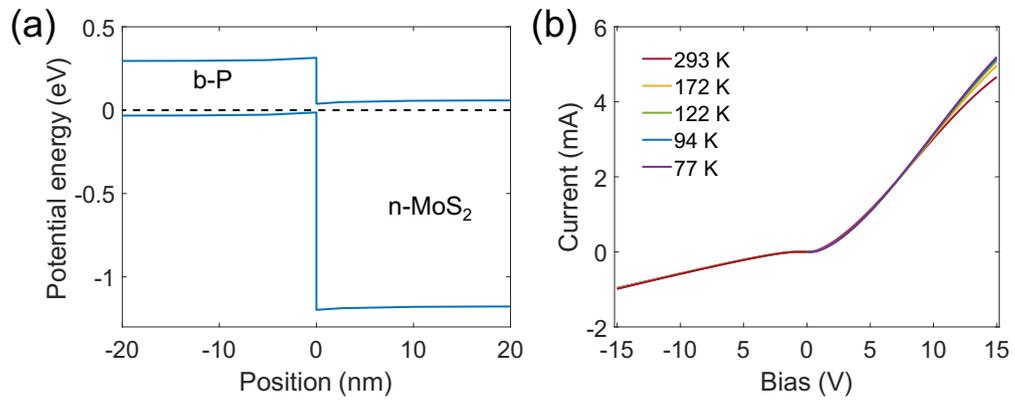

**Figure 4**: (a) Simulated equilibrium band diagram of the b-P/MoS$_2$ p-n junction (b) Simulated b-P LED current-voltage characteristics at different temperatures.

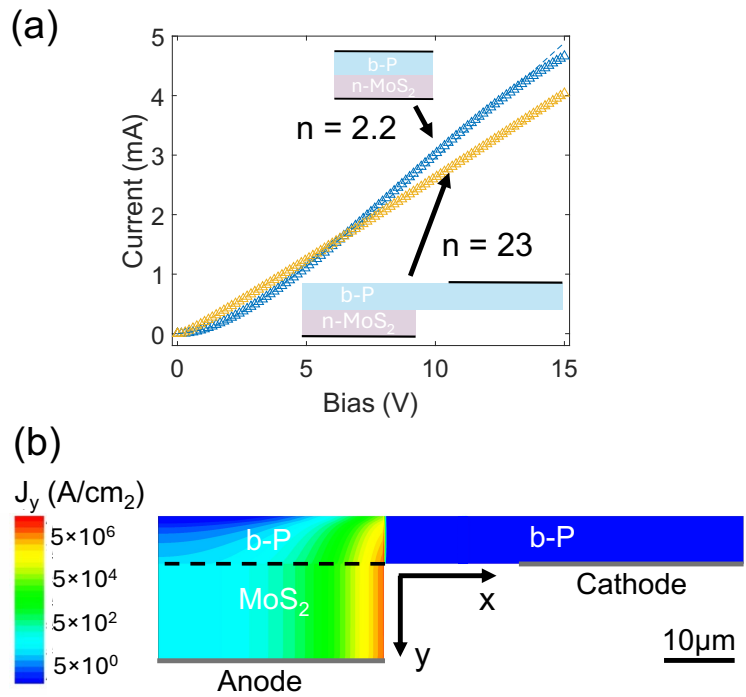

**Figure 5**: (a) Room-temperature forward-bias current-voltage characteristics calculated using the 1D and 2D models. The extracted ideality factors from the Shockley model fit are indicated. (b) The y-component of the total current density, Jy, in the 2D model at a bias of 15V at room temperature. The grey solid lines represent ohmic contacts in the model. Scale bar along the x-axis is 10 μm. Y-axis is scaled ×75 with respect to x-axis.

AUTHOR INFORMATION

**Corresponding Author**

*Stéphane Kéna-Cohen - Department of Engineering Physics, Polytechnique Montréal, H3T 1J4, Montreal, Canada

Email: s.kena-cohen@polymtl.ca

**Author Contributions**

J.B. and S.K.C. conceived the project and J.B. carried out the device fabrication, optoelectronic characterization and modelling. E.R. measured the TEM under the supervision of O.M and M.C.P helped setup the MIR EL imaging under the supervision of E.R. All authors contributed to the writing of the manuscript.

**Acknowledgements**

The authors would like to thank Professor Rémo A. Masut for a critical reading of the manuscript and for his valuable feedback. This work was funded by the Natural Sciences and Engineering Council of Canada Discovery Grant Program and the Canada Research Chair Program. We would also like to acknowledge design and fabrication support from CMC Microsystems and Canada's National Design Network (CNDN).

ASSOCIATED CONTENT

**Supporting Information**.

TEM measurement, PL and EL measurements, b-P LED flake thickness optimization using the classical oscillating electric dipole model, IV curves fit using the Shockley model, b-P resistor measurement, temperature-dependent PL spectra, finite-element modeling of the b-P LED current-voltage characteristics, external quantum efficiency calculation, MIR PL and EL maps of a b-P LED. (PDF)

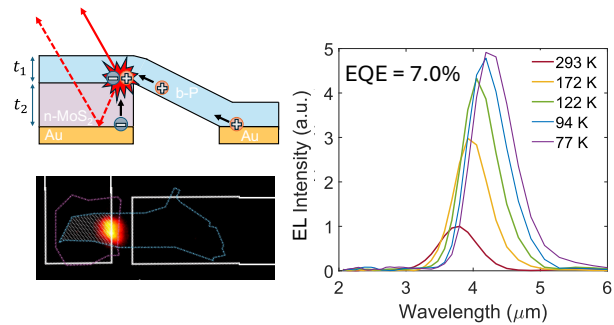

Figure 6: For Table of Contents Only

# Contents

## 1- Enhanced TEM image of second b-P/MoS$_2$

To verify the presence of an interface layer between b-P and MoS$_2$, we used TEM to image a second b-P (80 nm)/ MoS$_2$ (150 nm) junction fabricated under identical conditions and undamaged during the sample preparation. No oxide at the junction was detected during the EDS mapping. The brighter region of ~ 1 nm at the interface between b-P and MoS$_2$ consists of atomic layers with a slightly lower density. The layered structures of both b-P and MoS$_2$ are clearly visible.

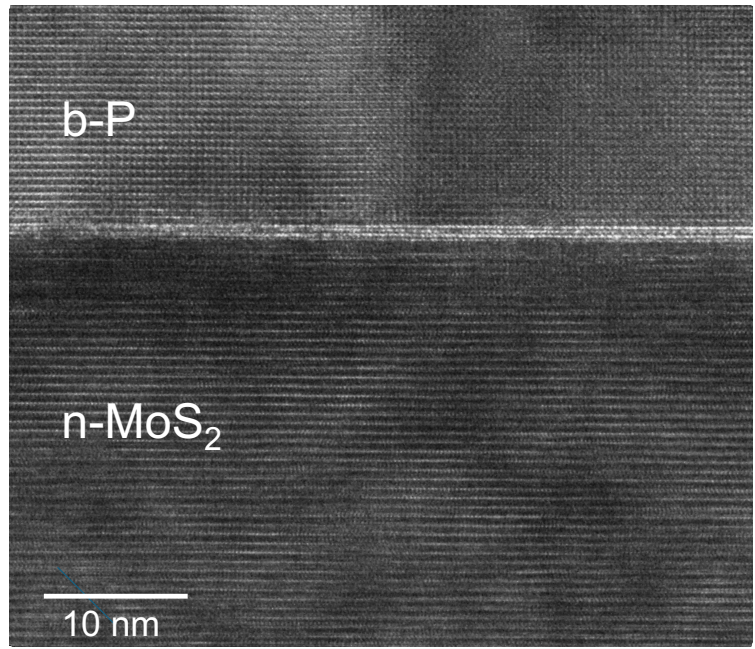

Figure S1: TEM image of an undamaged b-P/MoS$_2$ junction

## 2- Dipole radiation model for b-P LED flake thickness optimization

The Purcell factor $F$ and the outcoupling efficiency $\eta_{out}$ are calculated by modelling classical dipole radiation in an inhomogenous environment[1,2]. The refractive indices of b-P, MoS$_2$ and Au are taken from the literature[3,4,5]. For the calculation of the dipole dissipated power spectrum $\frac{dP}{du}$, the normalized in-plane wavenumber $u$ is defined as:

$$u = \frac{k_\parallel}{k_0 n_{bP}} \quad (1)$$

where $k_\parallel$ is the in-plane wavenumber, $k_0 = \frac{2\pi}{\lambda}$ is the vacuum wavenumber and $n_{bp}$ is the refractive index of b-P at the emission wavelength λ = 3.8 μm. For every normalized in-plane wavenumber $u$, the local electromagnetic field at the dipole position is calculated, the field being a sum of the dipole source term and the reflected field by the optical environnement of the dipole. The local electromagnectic field is then used to compute the density of dissipated dipole power $\frac{dP}{du}$ in every wavenumber $u$ (mode) of the optical structure. The Purcell factor $F$ is calculated as:

$$F = \int_0^\infty \frac{dP}{du}(u) du^2 = \int_0^\infty 2u \frac{dP}{du}(u) du \quad (2)$$

The outcoupling efficiency is defined as the fraction of dissipated power radiated in the viewing direction to the Purcell factor:

$$\eta_{out} = \frac{1}{F} \int_0^{1/n_{bp}} 2u \frac{dP_{out}}{du}(u) du \quad (3)$$

where $\frac{dP_{out}}{du}$ is the dipole power density radiated outside the struture. Note that for the calculation, we neglected the spectral distribition of b-P emission and that the calulation was performed only at λ = 3.8 μm since the optical structure is not resonnant.

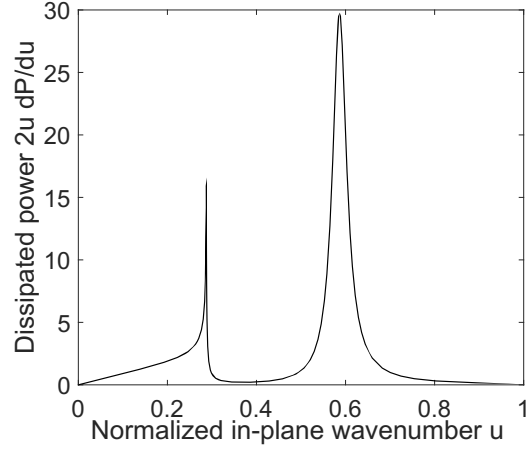

Figure S2: Dissipated dipole power spectrum $2udP/du$ as a function of the normalized in-plane wavenumber $u$ for the optimized LED structure with b-P and $MoS_2$ thickness of 75 nm and 155 nm respectively.

The outside angular emission pattern of the dipole can be calculated from the power dissipation spectrum:

$$P(\theta) = \frac{2}{n_{bP}^2} \frac{\cos(\theta)}{\pi} u \frac{dP}{du}(u) \qquad (4)$$

$$u(\theta) = \frac{k_\parallel(\theta)}{k_0 n_{bP}} = \frac{\sin(\theta)}{n_{bP}} \qquad (5)$$

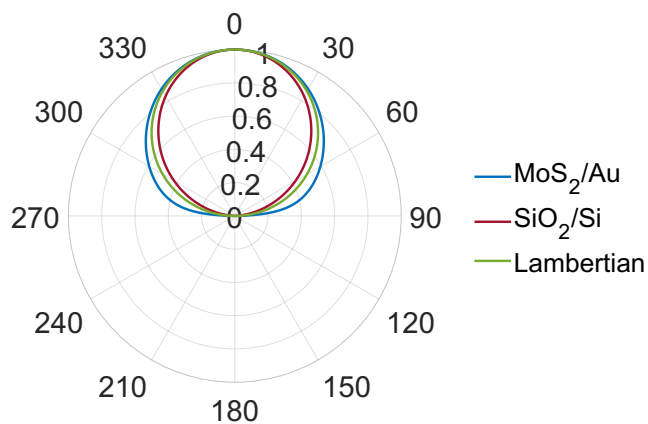

Figure S3: Calculated emission pattern for b-P on the optimized MoS$_2$/Au structure and over the SiO$_2$/Si region and compared to a perfect Lambertian emitter

### 3- PL image of a similar b-P/MoS$_2$ LED

A PL image of a similar b-P/MoS$_2$ LED was acquired using a MIR camera (Telops MS M350) and a CaF$_2$ tube lens with a 200 mm focal length. The MIR image was captured by exciting the b-P/MoS$_2$/Au active area using 808 nm CW laser diode at a 5 mW pump power. A second image was captured by exciting a region over the b-P/Au stack shown by the red circle in Figure S4(b) but no PL signal could be detected by the camera due to significant quenching of the b-P emission by the Au contact.

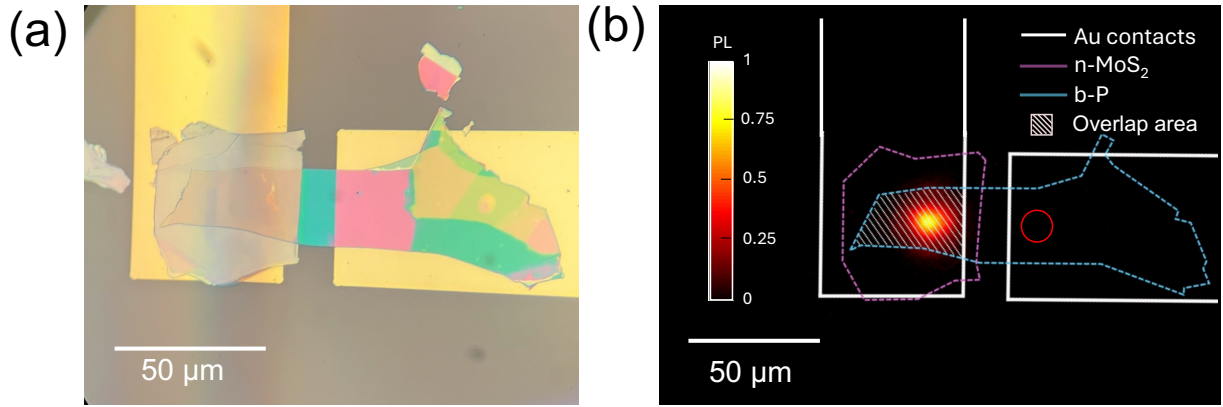

Figure S4: (a) Optical microscope image of a similar b-P/n-MoS$_2$ device. The b-P flake thickness are 80 nm and 150 nm respectively. (b) Relative PL intensity map taken with a MIR camera with the 808 nm laser exciting the b-P/MoS$_2$/Au active area and the b-P/Au region (red circle).

## 4- Experimental IV curves fit using the Shockley model

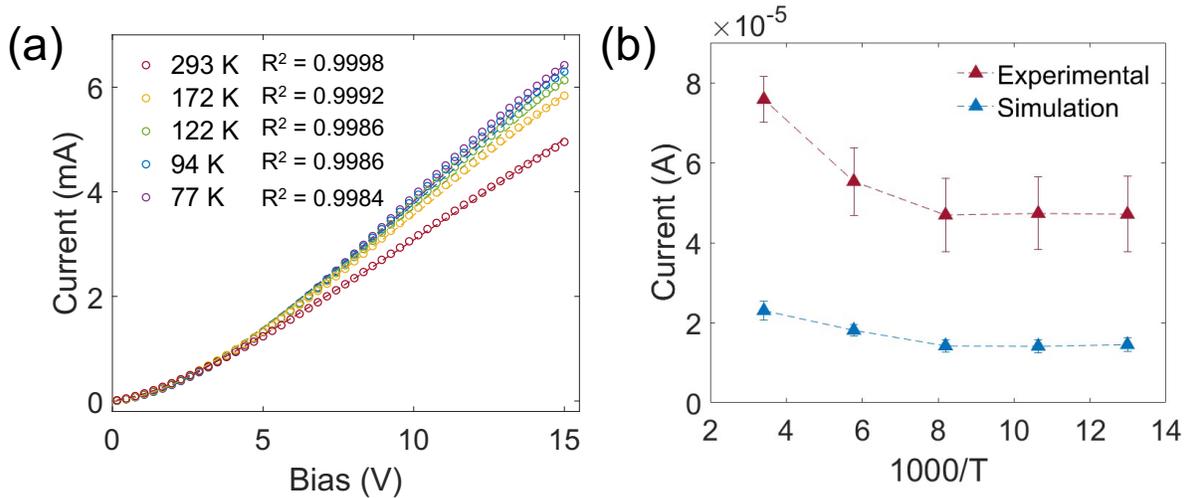

Figure S5: (a) Forward bias experimental IV curves fit with the Shockley model. (b) Fitted reverse saturation current of the Shockley model as a function of temperature for the experimental and simulated I-V curves.

## 5- b-P resistor measurement

The b-P doping concentration is estimated from the resistance value (R = 3.47 ± 0.03) kΩ is extracted from the IV of Figure S4.

$$N_A = \frac{L}{q\mu_p w t R} \tag{6}$$

The doping of b-P was estimated to be $N_A = (6 \pm 1) \times 10^{17}$ cm$^{-3}$ considering a zigzag hole mobility of $(400 \pm 20)$ cm$^2$V$^{-1}$s$^{-1}$ at room temperature [6], the device's length $L$, width $w$, thickness $t$ and assuming negligeable contact resistance. Note that both the fabricated LED and the resistor are conducting current along the zigzag direction. The orientation of the b-P flakes was confirmed using polarized PL measurements.

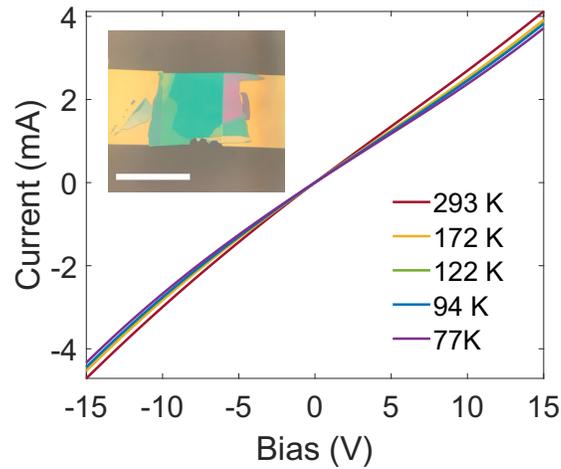

Figure S6: Temperature dependant current-voltage curve of a 70 nm thick b-P resistor. The resistor dimensions are 40 µm length by 45 µm width. Inset: Optical microscope image of the b-P resistor (scale bar 50 µm).

## 6- Temperature-dependent PL measurements

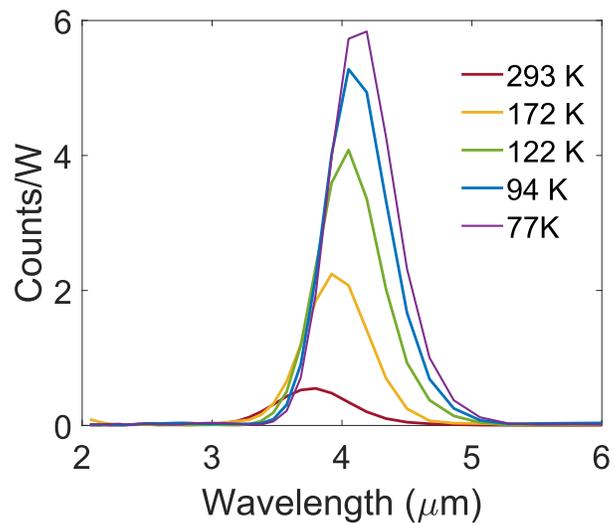

Figure S7: PL spectra of the LED b-P flake on the MoS$_2$/Au structure at various temperature. The spectra are obtained by pumping with a 808 nm CW laser diode at 3.2 mW.

## 7- Finite-element modeling of the b-P LED current-voltage characteristics

The Caughey-Thomas model[7] for the high-field velocity saturation of carriers was also included in the model with the following temperature dependance:

$$v_{sat} = A_{vsat} - B_{vsat}\left(\frac{T}{300K}\right) \qquad (7)$$

With A=1.7×10$^7$cm/s and B=3.6×10$^6$ cm/s. The radiative recombination coefficient was estimated using the Shockley-von-Roosbroeck relation[8] to B=9×10$^{-11}$ cm$^3$/s. The SRH and Auger coefficient were taken from the literature[8]. All other models used the default parameters provided by the Sentaurus device software.

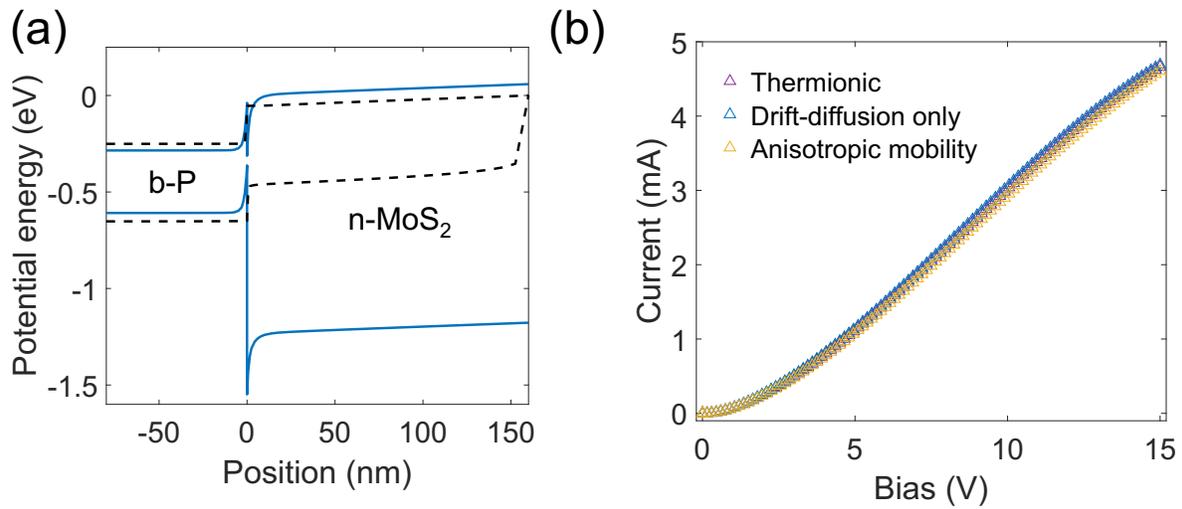

Figure S8: (a) Simulated band diagram of the b-P/MoS$_2$ p-n junction at a 5 V bias. The black dashed lines show the quasi-Fermi levels. (b) Simulated LED I-V characteristics of the 2D model with thermionic, drift-diffusion and anisotropic mobility models with $\mu_y = \frac{1}{2}\mu_x$.

Table S1: Fit parameters using the Shockley equation on the 1D finite element model at room temperature

| Fit parameter | Value |
|---|---|
| $I_s$ | $(5.6 \pm 1.5) \times 10^{-7}$ A |
| n | $2.2 \pm 0.1$ |
| $R_s$ | $(3641 \pm 12)$ Ω |

**8- EQE and PLQY rolloff comparison**

Figure S9 compares the room temperature EQE curve of Figure 3(b) as a function of the carrier generation rate assuming a uniform current injection through the device area of 550 um² corresponding to the b-P/MoS₂/Au overlap region (blue curve) and a uniform current injection through the active area of 120 um² estimated using our finite element simulation (yellow curve). The device PLQY as a function of carrier generation rate is also shown. The EQE carrier generation rate is calculated as:

$$GR_{EQE} = \frac{I}{q t_{bP} A} \quad (8)$$

where $I$ is the current, $q$ the elementary charge, $t_{bP}$ is the b-P flake thickness and $A$ is either the overlap or active area. The PLQY was obtained by pumping the b-P/MoS₂/Au active area with the 808 nm CW laser diode and varying the pump power. The PLQY carrier generation rate is calculated by assuming a uniform laser spot intensity on the sample and a uniform generation rate within the b-P flake thickness:

$$GR_{plqy} = \frac{N_p}{t_{bP} A_p} \alpha_{stack} \quad (9)$$

where $N_p$ is the number of pump photons incident on the device per second, $t_{bP}$ is the b-P flake thickness, $A_p$ is the pump area and $\alpha_{stack}$ is the absorption of the b-P/MoS₂/Au stack calculated using transfer matrix formalism.

Using the overlap area for the EQE leads to an Auger-induced roll-off at $10^{26}$ cm$^{-3}$s$^{-1}$. However, we see that the PLQY does not show any Auger recombination at those carrier densities. This is indicative that the EQE generation rate is underestimated by using the overlap area which is much larger than the active region of the device.

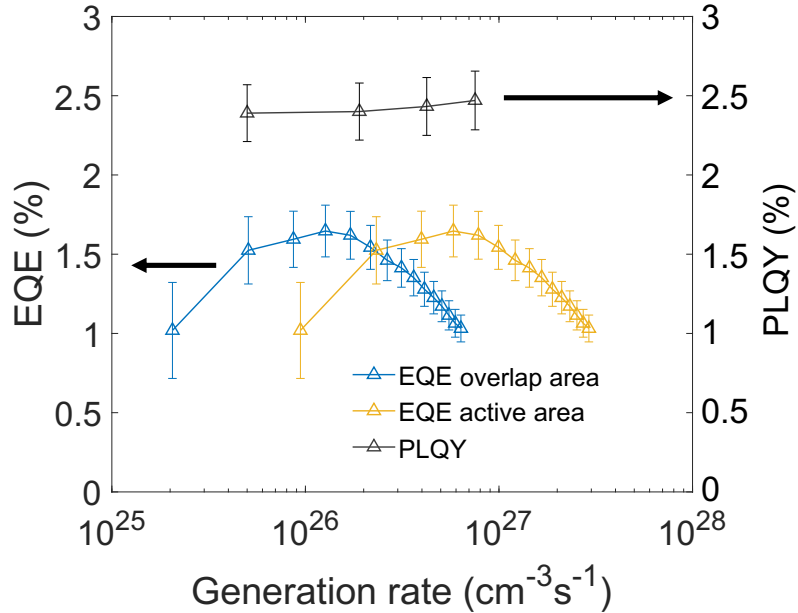

Figure S9: Comparison of the room temperature b-P LED PLQY and EQE as a function of the carrier generation rate assuming a uniform current injection through the overlap area (blue curve) and the active area (yellow curve).

### 9- EL mapping of a similar b-P LED

Similarly to the PL image of Figure S3, an EL intensity map was obtained by using a MIR camera (Telops MS M350) and a CaF$_2$ tube lens with a 200 mm focal length. The MIR image was captured while driving a 1.5 mA current in the device. The image clearly shows the EL emission being concentrated on the b-P flake edge close to the contact while no EL emission is observed on most of the device active area assumed to be the overlap region, confirming our simulation results.

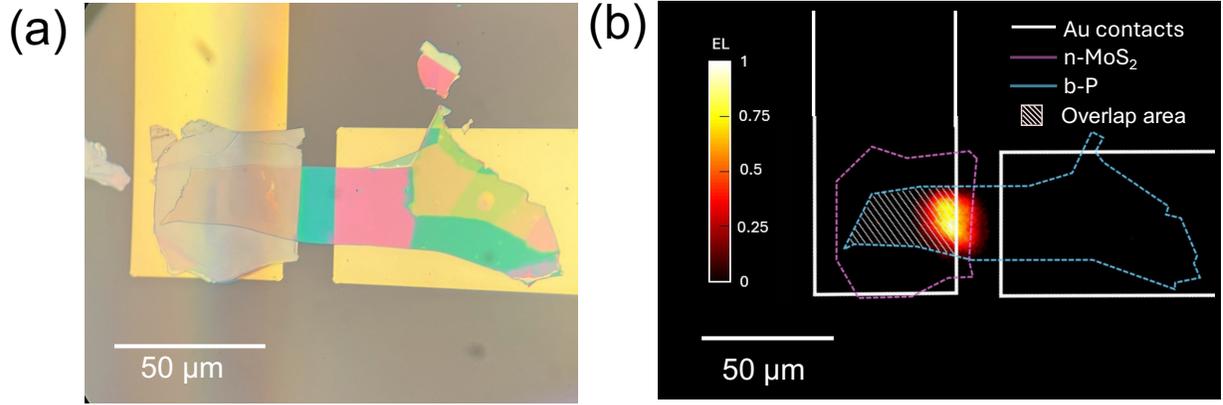

Figure S10: (a) Optical microscope image of a similar b-P/n-MoS$_2$ device. The b-P flake thickness are 80 nm and 150 nm respectively. (b) Relative EL intensity map taken with a MIR camera with while driving the device with a 1.5 mA current.

**10- EQE and PLQY calculation**

The absolute spectral responsivity $R(\lambda)$ of the InSb detector was calibrated using a 1550 nm CW laser diode with a know power entering the FTIR spectrometer and using the known relative spectral response of the detector. The EQE is calculated as follows:

$$\eta_{EQE} = \frac{q}{hcI\,T_{obj}} \int \frac{\lambda\, S_{EL}(\lambda)}{R(\lambda)} d\lambda \tag{10}$$

Where $S_{EL}$ is the measured EL spectra, $I$ is the current injected into the LED, $q$ is the elementary charge, $h$ is the Plank constant and $c$ is the speed of light and $T_{obj}$ is the transmission of the reflective objective. The transmission of the reflective objective was measured by scattering the 1550 nm CW laser diode on a Spectralon Diffuse Reflectance Standard with a known reflectivity to emulate a Lambertian emission. The power collected by the objective was then measured using a beam-splitter. We find that roughly 1% of the light emitted by a Lambertian source at 1550 nm is transmitted through the objective to the FTIR ($T_{obj} = 0.01$). We assume this value to be independent of wavelength. Similarly, the PLQY is calculated as:

$$\eta_{PLQY} = \frac{1}{\lambda_p \alpha_{stack} PT_{obj}} \int \frac{\lambda\, S_{PL}(\lambda)}{R(\lambda)} d\lambda \qquad (11)$$

where $\lambda_p$ is the pump wavelenght, $P$ is the pump power and $\alpha_{stack}$ is the absorption of the b-P/MoS2/Au stack calculated using transfer matrix formalism.

## 11- MIR LED EQE benchmark comparison

Table S1: Benchmark comparison of MIR LED EQE at room temperature

| Structure | Peak wavelength (μm) | Peak EQE (%) | Reference |
|---|---|---|---|
| b-P/n-MoS$_2$/Au | 3.8 | 1.6 | Our work |
| ITO/MoS$_2$/b-P/Al$_2$O$_3$/Au | 3.65 | 1.5 avg (4.43 best) | 9 |
| b-P/MoS$_2$ | 3.8 | 0.03 | 10 |
| graphite/b-P/graphite | 3.65 | 0.09 | 11 |
| HgCdTe with CdZnTe lens | 3.4 | 0.08 | 12 |
| HgCdTe microcavity | 3.2 | 0.02 | 13 |
| InAs/InAsSbP double heterostructure | 3.3 | 0.018 | 14 |
| InAs/InAsSb stained layer superlattice | 4.2 | 0.35 | 15 |
| InAs/GaInSb/InAs/AlGaAsSb quantum cascade | 3.3 | 0.03 | 16 |
| InAsSb/InAs quantum cascade | 3.5 | 0.003 | 17 |
| InAs/GaInSb/InAs interband cascade | 3.8 | 0.65 | 18 |
| $Al_2O_3$/ITO/HgTE colloïdal quantum dots/Au | 4.0 | 0.1 | 19 |
| BostonElectronic LED38TO8TEC | 3.8 | 0.1 | 10 |
| Hamamatsu L13454 | 3.9 | 0.9 | 10 |